\documentclass[aps,prd,twocolumn,showpacs,superscriptaddress,groupedaddress]{revtex4}
\usepackage{graphicx}
\usepackage{dcolumn}
\usepackage{bm}
\usepackage{amssymb}
\usepackage{url}
\usepackage{soul}
\usepackage{multirow}
\usepackage{color} 
\usepackage{ulem}

\begin{document}

\hyphenation{state its}

\title{Search for the process $e^+e^-\to J/\psi + X(1835)$
at $\sqrt{s}\approx10.6$~GeV}

\noaffiliation
\affiliation{University of the Basque Country UPV/EHU, 48080 Bilbao}
\affiliation{Beihang University, Beijing 100191}
\affiliation{Budker Institute of Nuclear Physics SB RAS and Novosibirsk State University, Novosibirsk 630090}
\affiliation{Faculty of Mathematics and Physics, Charles University, 121 16 Prague}
\affiliation{University of Cincinnati, Cincinnati, Ohio 45221}
\affiliation{Deutsches Elektronen--Synchrotron, 22607 Hamburg}
\affiliation{Hanyang University, Seoul 133-791}
\affiliation{University of Hawaii, Honolulu, Hawaii 96822}
\affiliation{High Energy Accelerator Research Organization (KEK), Tsukuba 305-0801}
\affiliation{IKERBASQUE, Basque Foundation for Science, 48011 Bilbao}
\affiliation{Indian Institute of Technology Guwahati, Assam 781039}
\affiliation{Indian Institute of Technology Madras, Chennai 600036}
\affiliation{Institute of High Energy Physics, Chinese Academy of Sciences, Beijing 100049}
\affiliation{Institute of High Energy Physics, Vienna 1050}
\affiliation{Institute for High Energy Physics, Protvino 142281}
\affiliation{INFN - Sezione di Torino, 10125 Torino}
\affiliation{Institute for Theoretical and Experimental Physics, Moscow 117218}
\affiliation{J. Stefan Institute, 1000 Ljubljana}
\affiliation{Kanagawa University, Yokohama 221-8686}
\affiliation{Institut f\"ur Experimentelle Kernphysik, Karlsruher Institut f\"ur Technologie, 76131 Karlsruhe}
\affiliation{Korea Institute of Science and Technology Information, Daejeon 305-806}
\affiliation{Korea University, Seoul 136-713}
\affiliation{Kyungpook National University, Daegu 702-701}
\affiliation{\'Ecole Polytechnique F\'ed\'erale de Lausanne (EPFL), Lausanne 1015}
\affiliation{Faculty of Mathematics and Physics, University of Ljubljana, 1000 Ljubljana}
\affiliation{Max-Planck-Institut f\"ur Physik, 80805 M\"unchen}
\affiliation{School of Physics, University of Melbourne, Victoria 3010}
\affiliation{Moscow Physical Engineering Institute, Moscow 115409}
\affiliation{Moscow Institute of Physics and Technology, Moscow Region 141700}
\affiliation{Graduate School of Science, Nagoya University, Nagoya 464-8602}
\affiliation{Nara Women's University, Nara 630-8506}
\affiliation{National Central University, Chung-li 32054}
\affiliation{National United University, Miao Li 36003}
\affiliation{Department of Physics, National Taiwan University, Taipei 10617}
\affiliation{H. Niewodniczanski Institute of Nuclear Physics, Krakow 31-342}
\affiliation{Nippon Dental University, Niigata 951-8580}
\affiliation{Niigata University, Niigata 950-2181}
\affiliation{University of Nova Gorica, 5000 Nova Gorica}
\affiliation{Osaka City University, Osaka 558-8585}
\affiliation{Pacific Northwest National Laboratory, Richland, Washington 99352}
\affiliation{Panjab University, Chandigarh 160014}
\affiliation{Peking University, Beijing 100871}
\affiliation{University of Pittsburgh, Pittsburgh, Pennsylvania 15260}
\affiliation{Research Center for Electron Photon Science, Tohoku University, Sendai 980-8578}
\affiliation{University of Science and Technology of China, Hefei 230026}
\affiliation{Seoul National University, Seoul 151-742}
\affiliation{Soongsil University, Seoul 156-743}
\affiliation{Sungkyunkwan University, Suwon 440-746}
\affiliation{School of Physics, University of Sydney, NSW 2006}
\affiliation{Tata Institute of Fundamental Research, Mumbai 400005}
\affiliation{Excellence Cluster Universe, Technische Universit\"at M\"unchen, 85748 Garching}
\affiliation{Toho University, Funabashi 274-8510}
\affiliation{Tohoku Gakuin University, Tagajo 985-8537}
\affiliation{Tohoku University, Sendai 980-8578}
\affiliation{Department of Physics, University of Tokyo, Tokyo 113-0033}
\affiliation{Tokyo Institute of Technology, Tokyo 152-8550}
\affiliation{Tokyo Metropolitan University, Tokyo 192-0397}
\affiliation{Tokyo University of Agriculture and Technology, Tokyo 184-8588}
\affiliation{University of Torino, 10124 Torino}
\affiliation{CNP, Virginia Polytechnic Institute and State University, Blacksburg, Virginia 24061}
\affiliation{Wayne State University, Detroit, Michigan 48202}
\affiliation{Yamagata University, Yamagata 990-8560}
\affiliation{Yonsei University, Seoul 120-749}
  \author{X.~H.~He}\affiliation{Peking University, Beijing 100871} 
  \author{J.~Wang}\affiliation{Peking University, Beijing 100871} 
  \author{Y.~Ban}\affiliation{Peking University, Beijing 100871} 
  \author{P.~Wang}\affiliation{Institute of High Energy Physics, Chinese Academy of Sciences, Beijing 100049} 
  \author{I.~Adachi}\affiliation{High Energy Accelerator Research Organization (KEK), Tsukuba 305-0801} 
  \author{H.~Aihara}\affiliation{Department of Physics, University of Tokyo, Tokyo 113-0033} 
  \author{D.~M.~Asner}\affiliation{Pacific Northwest National Laboratory, Richland, Washington 99352} 
  \author{V.~Aulchenko}\affiliation{Budker Institute of Nuclear Physics SB RAS and Novosibirsk State University, Novosibirsk 630090} 
  \author{T.~Aushev}\affiliation{Institute for Theoretical and Experimental Physics, Moscow 117218} 
  \author{A.~M.~Bakich}\affiliation{School of Physics, University of Sydney, NSW 2006} 
% ...
\collaboration{The Belle Collaboration}

\date{\today}

\begin{abstract}
We report the results of a search for the $X(1835)$ state in
the process $e^+e^-\to J/\psi+X(1835)$ using a data sample of 672 fb$^{-1}$
collected with the Belle detector at and near the $\Upsilon(4S)$ resonance
at the KEKB asymmetric-energy $e^+e^-$ collider.
No significant evidence is found for this process, and an
upper limit is set on its cross section times the branching fraction:
$\sigma_{\rm Born}(e^+e^- \to J/\psi X(1835)) \cdot$
{${\cal B}(X(1835)\to \ge 3$ charged tracks)} $< 1.3 \ {\rm fb}$ at 90\%
confidence level.
\end{abstract}

\maketitle

\begin{figure}
\includegraphics[scale=0.6, bb=0 5 300 200, clip]{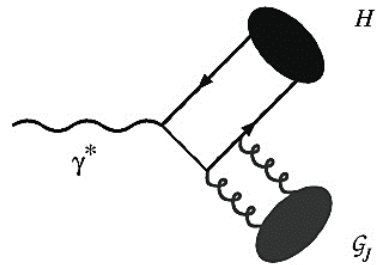}
\caption{\label{feyn} Possible Feynman diagram for $\gamma^* \to
H+\mathcal{G}_J$~\cite{Brodsky}.}
\end{figure}

The BESII
Collaboration observed a resonance, the $X(1835)\to\pi^+\pi^-\eta^{\prime}$, in the radiative decay $J/\psi\to\gamma\pi^+\pi^-\eta^{\prime}$,
with a 7.7$\sigma$ statistical significance~\cite{Ablikim}.
Recently, the structure has been confirmed by BESIII
in the same process with a statistical significance greater than 20$\sigma$~\cite{Ablikim_2011}.
From a fit with a Breit-Wigner function, the mass and width are determined
to be $1836.5\pm3.0(\rm stat.)^{+5.6}_{-2.1}(\rm syst.)$ MeV/$c^2$ and
$190\pm9(\rm stat.)^{+38}_{-36}(\rm syst.)$ MeV, respectively, with a
product branching fraction of
${\cal B}(J/\psi\to\gamma X)\cdot {\cal B}(X\to \pi^+\pi^-\eta')
=[2.87\pm0.09(\rm stat.)^{+0.49}_{-0.52}(\rm syst.)]\times 10^{-4}$\cite{Ablikim_2011}.
The Belle Collaboration also searched for the $X(1835)$ in two-photon
collisions, but no strong evidence was found~\cite{twophoton}.
Many theoretical models have been proposed to interpret its underlying structure.
Some consider the $X(1835)$ as a radial excitation of the $\eta^{\prime}$ ~\cite{Huang, Klempt};
a $p\bar{p}$ bound state ~\cite{Loiseau, Datta, SLZhu}; a glueball
candidate ~\cite{Kochelev, XGHe, BALi, Hao}; or
a $\eta_{c}$-glueball mixture ~\cite{Kochelev1}.
$C$-even glueballs can be studied in the process
$e^+e^- \to\gamma^{*} \to H+\mathcal{G}_J$~\cite{Brodsky},
where $H$ denotes a $c\bar{c}$ quark pair or charmonium state
and $\mathcal{G}_J$ is a glueball, as shown in Fig.~\ref{feyn}.
In this paper, we search for $X(1835)$ in the process
$e^+e^-\to J/\psi X(1835)$ at $\sqrt{s}\approx10.6$~GeV.

This analysis uses a $604$ fb$^{-1}$ data sample
collected with the Belle detector~\cite{belle_detector} at the $\Upsilon(4S)$
resonance and $68$ fb$^{-1}$ 60 MeV below it  at the KEKB
asymmetric-energy $e^+e^-$ collider~\cite{kekb}.
The Belle detector is a large-solid-angle magnetic spectrometer
that consists of a
silicon vertex detector (SVD), a 50-layer central drift
chamber (CDC), an array of aerogel threshold Cherenkov counters (ACC),
a barrel-like arrangement of time-of-flight scintillation counters
(TOF), and an electromagnetic calorimeter (ECL) comprised of
CsI(Tl) crystals located inside a superconducting solenoid coil that
provides a 1.5T magnetic field. An iron flux return located outside
of the coil is instrumented to detect $K^0_L$ mesons and to identify
muons (KLM). Two different inner detector
configurations were used: a 2.0 cm radius beam pipe and a 3-layer
silicon vertex detector for the first $155$ fb$^{-1}$ data, and a 1.5 cm
radius beam pipe with a 4-layer vertex detector for the remaining
data sample.

A Monte Carlo (MC) simulation based on the {\sc babayaga} event
generator~\cite{mc}, in which the initial state radiation (ISR) correction is
taken into account, is used to estimate the selection efficiency.
We assume the minimum remaining system energy (after initial state radiation)
to be 8 GeV.
To incorporate the $X(1835)$ $J/\psi$ reaction into {\sc babayaga},
the two-body final state is assumed to be distributed according to
$1+\cos^2\theta$ in the $e^+e^-$ center-of-mass (CM) system,
where $\theta$ is the angle between the $J/\psi$ and $e^-$
beam direction in the CM system.
The mass of $X(1835)$
is generated according to a Breit-Wigner function, with the reported
mass of 1836 MeV/$c^2$ and width of 190 MeV.
The efficiency is calculated using $e^+e^-\to J/\psi
X(1835)(\gamma)$ signal events, where the $J/\psi$ decays to $e^+e^-$ or
$\mu^+\mu^-$ and the $X(1835)$ decays to $\eta'$$\pi^+$$\pi^-$, followed by
$\eta' \to \eta$$\pi^+$$\pi^-$ and $\eta \to \gamma\gamma$.

The $J/\psi$ reconstruction procedure is similar to that described
in Ref.~\cite{Abe}.
Oppositely charged tracks that are both identified either as muons
or electrons are combined as a $J/\psi$ candidate.
To correct for final state radiation and bremsstrahlung, photons within 50 mrad of
the $e^\pm$ are included in the $e^+e^-$ invariant mass calculation.
The lepton identification efficiencies are $96\%$ and
$98\%$ for $\mu^{\pm}$ and $e^{\pm}$, respectively.
The two lepton candidate tracks are required to have a common
vertex, with a distance to the IP in the $r\phi$ plane (transverse to the beam direction) smaller
than 100 $\mu$m. The $J/\psi$ signal region is defined by the mass window
$\vert{M_{l^+l^-}-M_{J/\psi}}\vert < 30$ MeV/$c^2$ ($\sim
2.5\sigma$), common for both dimuon and dielectron channels. We
also define a sideband region as 70
MeV/$c^2<\vert{M_{l^+l^-}-M_{J/\psi}}\vert <190$ MeV/$c^2$ , which
is used to estimate the contribution from the dilepton combinatorial
background under the $J/\psi$ peak. A mass-constrained fit to the
reconstructed $J/\psi$ candidates is then performed to improve
their momentum resolution. The mass of the system recoiling against
a reconstructed $J/\psi$ is determined from:
\begin{equation}
M_{\rm recoil}=\sqrt{(E_{CM}-E^*_{J/\psi})^2-p_{J/\psi}^{*2}},
\end{equation}
where $E_{CM}$ is the CM energy of $e^+e^-$ collisions,
and $E^*_{J/\psi}$ and $p_{J/\psi}^{*}$
are the energy and momentum of the $J/\psi$ candidate in the CM system,
respectively.

\begin{figure}
\includegraphics[scale=0.44]{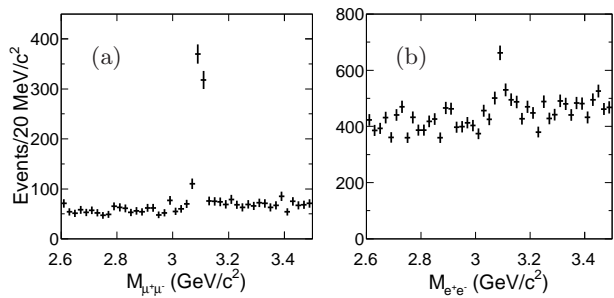}
\put(-215,90){(a)}
\put(-100,90){(b)}
\caption{\label{jpsi} Mass distribution for the $J/\psi$ candidates
reconstructed from $\mu^+\mu^-$(a) and $e^+e^-$(b)
in the region 0$<M_{\rm recoil}<$3 GeV/$c^2$ .
}
\end{figure}

The background due to initial state radiation with a hard photon
[radiative return to $J/\psi$($\psi(2S)$)]~\cite{Benayoun} and the QED
process $J/\psi e^+e^-$\cite{Chang2} is large. According to a
study reported in Ref.~\cite{Abe}, these backgrounds contribute
mainly to $N_{\rm ch}=3$ and $N_{\rm ch}=4$ events (where $N_{\rm ch}$ is the
number of charged tracks in an event). We suppress these
backgrounds by requiring $N_{\rm ch}>4$.
The mass distributions for $J/\psi$ candidates in the region
0$<M_{\rm recoil}<$3 GeV/$c^2$
after the selection are shown in Fig.~\ref{jpsi}.

\begin{figure}
\includegraphics[scale=0.44]{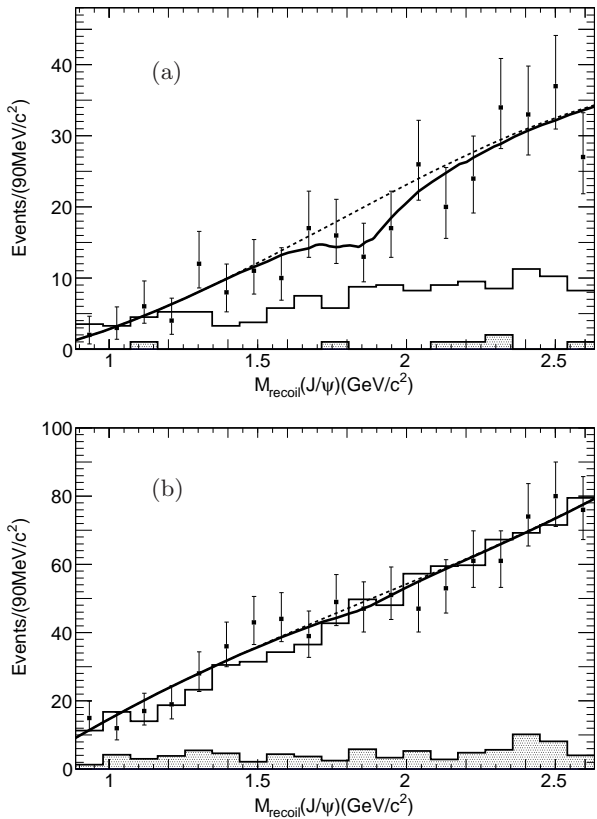}
\put(-190,288){(a)}
\put(-190,130){(b)}
\caption{\label{fitx}
Distribution of the recoil mass against the $J/\psi$ reconstructed from $\mu^+\mu^-$(a) and $e^+e^-$(b).
The points are data, the solid histograms represent the backgrounds
from the $J/\psi$ sideband, and the hatched histograms represent the
charmed- plus $uds$-quark backgrounds.
The solid lines are results of the fits in the low
recoil mass region and the dashed lines are the total background.
}
\end{figure}

The $M_{\rm recoil}$ distributions
are shown in Fig.~\ref{fitx}. The remaining backgrounds are mainly from two
sources.
One is the combinatorial dilepton events in the $J/\psi$ mass window that are estimated from the $J/\psi$ sideband data, as shown in Fig.~\ref{fitx}.
The other background is the non-prompt $J/\psi$ decay products from
excited charmonium states (such as $\psi'$, $\chi_{cJ}$). This
is found to contribute negligibly to the $J/\psi$ signal.
To understand the background from $\psi' \to \pi^+\pi^- J/\psi$ decays,
we reconstruct such events by combining the
detected $J/\psi$ mesons with any pair of oppositely charged pion tracks
and find fewer than five events
in the region $M_{\rm recoil} <$ 3 GeV/$c^2$  at 95\% C.L.
$J/\psi$ mesons from $B$ decay are kinematically forbidden
to produce a recoil mass below 3 GeV/$c^2$.

In order to understand the $J/\psi$ peaking background, we analyze
a sample of continuum MC events at the $\Upsilon(4S)$ generated
with EvtGen~\cite{evtgen}.
After the selection criteria are applied, the surviving
background is less than the combinatorial lepton pair background, as shown in
Fig.~\ref{fitx}. Annihilation of two virtual photons in the process
$e^+e^- \to \gamma^*\gamma^* \to J/\psi\gamma^* \to J/\psi f\bar{f}$ may
contribute significantly to the background in the low $M_{\rm recoil}$ mass
region, where $f\bar{f}$ denotes a pair of light quarks hadronizing into multi-hadrons.
This type of background is suppressed by the $N_{\rm ch}>4$ cut.

We search for an $X(1835)$ signal using an unbinned maximum likelihood
fit to the $M_{\rm recoil}$ distributions shown in Fig.~\ref{fitx},
in the region 0.8 GeV/c$^2<M_{\rm recoil}<2.8$ GeV/$c^2$.
The signal shape is fixed to the MC simulation using the mass and width
from the BESIII measurement~\cite{Ablikim_2011}.
The background is represented by a third-order Chebychev function.
A simultaneous fit is performed for the $\mu^+\mu^-$ and $e^+e^-$ channels,
which constrains the expected signal from $J/\psi \to \mu^+\mu^-$
and $J/\psi \to e^+e^-$ to be consistent with the ratio of
$\varepsilon_i$ and ${\cal B}_i$,
where $\varepsilon_i$ and ${\cal B}_i$ are the efficiency and branching
fraction for the two channels, respectively.
The $\varepsilon_i$ values are obtained from MC simulation including ISR.
The results of the fit are shown in Table~\ref{fit} and Fig.~\ref{fitx}.

\begin{table}
\caption{\label{fit}Fit results for the $M_{\rm recoil}$ region 0.8--
2.8 GeV/$c^2$}
\begin{ruledtabular}
\begin{tabular}{cccc}
{Mode}                                  & {$N_{\rm signal}$}  &
{$N_{\rm background}$}\\
\hline
$J/\psi \to {\mu^+\mu^-}$ &{$-20.0\pm20.0$} &{$340.0\pm18.0$}\\
$J/\psi \to {e^+e^-}$     &{$-7.5\pm7.6$}   &{$859.5\pm29.2$}\\
\end{tabular}
\end{ruledtabular}
\end{table}

The Born cross section is determined by the following formula derived from the second-order calculation of the
perturbation theory~\cite{born sec}:
\begin{equation}
\sigma_{\rm Born}=\sigma_{\rm measured}(\rm{non\textendash ISR})/\xi_{\rm Born},
\end{equation}
where $\sigma_{\rm measured}\rm{(non\textendash ISR)}$ is the cross section
when the energy of a radiative photon is less than 10 MeV. The value
of this cut-off energy $E_{\rm{rad.}\gamma}$ is arbitrary; the final
result is independent of this choice.
The factor $\xi_{\rm Born}$ relates the
measured cross section with radiative photons below the cut-off energy
to the Born cross section. From the QED calculation~\cite{born sec},
$\xi_{\rm Born}$ is determined to be 0.629 for $E_{\rm{rad.}\gamma}$ = 10 MeV.
The final Born cross section is then estimated as
\begin{equation}
\sigma_{\rm Born}=\frac{R_{\varepsilon} f_{\rm non\_isr}}{\xi_{\rm Born}}\times \frac{N_{\rm fit}}{
{\cal{L}}_{\rm int} \varepsilon_{\rm sum} {\cal{B}}_{\rm sum} },
\label{eq6}
\end{equation}
where $N_{\rm fit}$ is the sum of the fitted event yields in
the ${\mu^+\mu^-}$ and ${e^+e^-}$ modes,
the factor $R_{\varepsilon}$
is the ratio of the full and non$\textendash$ISR reconstruction efficiencies
and $f_{\rm non\_isr}$ is the fraction of non$\textendash$ISR events
depending on the final states that are incorporated using the signal MC
sample. For $E_{\rm{rad.}\gamma} = 10$ MeV, this part of the soft ISR
process accounts for approximately 65\% of the total. Here, ${\cal{L}}_{\rm int}$
is the integrated luminosity,
$\varepsilon_{\rm sum}$ is the total detection efficiency and
${\cal{B}}_{\rm sum}$ is the total branching fraction
of $J/\psi \to {\mu^+\mu^-}$ and ${e^+e^-}$ decays.

\begin{table}[ht]
\caption{\label{systematic}Contributions to the systematic uncertainties}
\begin{ruledtabular}
\begin{tabular}
{@{\hspace{0.5cm}}l@{\hspace{0.5cm}}@{\hspace{0.5cm}}c@{\hspace{0.5cm}}@{\hspace{0.5cm}}c@{\hspace{0.5cm}}}
\multirow{2}{*}{Source} &  \multicolumn{2} {c} {Syst. uncertainties (\%)} \\
                        & ${\mu^+\mu^-}$ & ${e^+e^-}$ \\
\hline
$Q^{2}$ dependence     & $5$   & $5$ \\
BG estimation          & $9$   & $13$\\
$J/\psi$ polarization  & $8$   & $10$ \\
$X(1835)$ width          & $16$  & $16$ \\
Track reconstruction   & $2$   & $2$ \\
Lepton identification  & $3$   & $3$ \\
MC statistics          & $3$   & $3$\\
\hline
Sum in quadrature      & $21$  &$24$ \\
\end{tabular}
\end{ruledtabular}
\end{table}

Since the fit does not return any significant signal in the $X(1835)$
mass region,
we set an upper limit on its production rate.
The upper limit of $\sigma_{\rm Born}$ is calculated by replacing
$N_{\rm fit}$ with the upper limit
on the signal yield at 90\% C.L. in Eq.~\ref{eq6}.
We integrate the likelihood function starting at
$N_{\rm event}=0$; the upper limit is set when the integral reaches 90\% of the total area.
The total  upper limit of $X(1835)$ events in the two
$J/\psi$ decay modes is $N_{\rm event}=46.7$ at 90\% C.L.

Systematic uncertainties listed in Table~\ref{systematic} are dominated by the following sources.
The form-factor dependence on $Q^2$ affects the shape
of the ISR tail. We replace the $1/Q^2$ dependence with $1/Q^4$
and find the results change by $5\%$; this is taken as the corresponding
systematic uncertainty. We change the minimum remaining system
energy (after ISR) from 8 GeV to 9 GeV to estimate the systematic uncertainty
from MC simulation.
The uncertainty from the background estimation is evaluated by the
variations in the result arising from changes in  the fitting range and background shape (the latter being obtained from fitting $M_{\rm recoil}$ on $J/\psi$ sideband data); fitting $M_{\rm recoil}$ including the signal region (1.7
GeV/$c^2$--2.2 GeV/$c^2$); and floating the background
parameters.
The quantum numbers $J^{PC}$ of the $X(1835)$ reported by BES are $0^{-+}$,
corresponding to a $(1+\cos^2\theta)$ polar angular distribution. We generate events with flat and $\sin^2\theta$ distributions to compare and
estimate the systematic uncertainty associated with different possible
polarizations of the $J/\psi$. The width of the $X(1835)$ remeasured by
BES\uppercase\expandafter{\romannumeral3} is $\Gamma$ = 190$\pm$9
$(\rm stat.)^{+38}_{-36}(\rm syst.)$ MeV~\cite{Ablikim_2011}; the
systematic uncertainty caused by different widths is
taken into account.

Other systematic uncertainties come from MC statistics (3\%), track
reconstruction efficiency (1\% per track) and lepton
identification uncertainty (1.5\% per lepton) in $J/\psi$ reconstruction.
The luminosity and branching ratio uncertainties are negligible.

The systematic uncertainties caused by the $J/\psi$ polarization for the two
decay modes $J/\psi \to {\mu^+\mu^-}$ and $J/\psi \to {e^+e^-}$
are correlated, which will expand or shrink
the likelihood functions in the same way. Other sources of
systematic uncertainties for the two $J/\psi$ decay modes are
uncorrelated. In the combination of the two $J/\psi$ decay modes,
some systematic uncertainties cancel. However, in the upper limit
calculation, we use just the systematic uncertainty for $J/\psi \to {e^+e^-}$,
which gives the most conservative result.

Since the recoil mass method is used in the analysis, the efficiency
of the $X(1835)$ selection  always coincides with the efficiency of $J/\psi$ reconstruction independently of the decay modes of $X(1835)$.
The MC simulation $e^+e^- \to J/\psi X(1835)$, where $X(1835)$ decays to $\eta'\pi^+\pi^-$ with
$\eta' \to \eta\pi^+\pi^-$, $\eta \to \gamma\gamma$,
is a mode with fewest
charged tracks that satisfies $N_{\rm ch}
> 4$ and thus has the lowest efficiency. Using this efficiency
in the upper limit calculation also gives a less restrictive upper
limit.

After taking into account the systematic uncertainty, the upper limit on $\sigma_{\it{born}}$ is 1.3 fb.

In summary, using a 672 fb$^{-1}$ data sample
collected with the Belle detector, we search for the $X(1835)$ state
by analyzing the $J/\psi$ recoil mass distribution from the assumed
process $e^+e^-\to J/\psi X(1835)$. No significant evidence
for $X(1835)$ production in this process is found.
An upper limit is set to
be: $\sigma_{\rm Born}(e^+e^- \to J/\psi X(1835)) \cdot$
{${\cal B}(X(1835)\to \ge 3$ charged tracks)} $<$ $1.3 \ {\rm fb}$  at $90\%$ C.L, including systematic uncertainties.
This upper limit is three orders of magnitude smaller than the cross section for
prompt production of the $J/\psi$ meson ~\cite{Abe}.
No evidence is found to support
the hypothesis of the $X(1835)$ as a glueball  produced in association with a $J/\psi$ in the Belle experiment.

We thank the KEKB group for the excellent operation of the
accelerator; the KEK cryogenics group for the efficient
operation of the solenoid; and the KEK computer group,
the National Institute of Informatics, and the
PNNL/EMSL computing group for valuable computing
and SINET4 network support.  We acknowledge support from
the Ministry of Education, Culture, Sports, Science, and
Technology (MEXT) of Japan, the Japan Society for the
Promotion of Science (JSPS), and the Tau-Lepton Physics
Research Center of Nagoya University;
the Australian Research Council and the Australian
Department of Industry, Innovation, Science and Research;
Austrian Science Fund under Grant No. P 22742-N16;
the National Natural Science Foundation of China under contract
No.~10575109, 10775142, 10825524, 10875115, 10935008 and 11175187;
the Ministry of Education, Youth and Sports of the Czech
Republic under contract No.~MSM0021620859;
the Carl Zeiss Foundation, the Deutsche Forschungsgemeinschaft
and the VolkswagenStiftung;
the Department of Science and Technology of India;
the Istituto Nazionale di Fisica Nucleare of Italy;
The WCU program of the Ministry Education Science and
Technology, National Research Foundation of Korea Grant No.
2011-0029457, 2012-0008143, 2012R1A1A2008330, 2013R1A1A3007772,
BRL program under NRF Grant No. KRF-2011-0020333, KRF-2011-0021196, BK21 Plus program,
and GSDC of the Korea Institute of Science and Technology Information;
the Polish Ministry of Science and Higher Education and
the National Science Center;
the Ministry of Education and Science of the Russian
Federation and the Russian Federal Agency for Atomic Energy;
the Slovenian Research Agency;
the Basque Foundation for Science (IKERBASQUE) and the UPV/EHU under
program UFI 11/55;
the Swiss National Science Foundation; the National Science Council
and the Ministry of Education of Taiwan; and the U.S.\
Department of Energy and the National Science Foundation.
This work is supported by a Grant-in-Aid from MEXT for
Science Research in a Priority Area (``New Development of
Flavor Physics''), and from JSPS for Creative Scientific
Research (``Evolution of Tau-lepton Physics'').

\end{document}